# Electromagnetic Force and Momentum

Masud Mansuripur, College of Optical Sciences
The University of Arizona, Tucson

**Status**. At the foundation of the classical theory of electrodynamics lie Maxwell's macroscopic equations, namely,

$$\nabla \cdot D(r,t) = \rho_{\text{free}}(r,t). \quad (1)$$

$$\nabla \times H(r,t) = J_{\text{free}}(r,t) + \partial D(r,t)/\partial t. \quad (2)$$

$$\nabla \times E(r,t) = -\partial B(r,t)/\partial t. \quad (3)$$

$$\nabla \cdot B(r,t) = 0. \quad (4)$$

In the above equations, $\rho_{\text{free}}$ and $J_{\text{free}}$ are densities of free charge and free current, displacement $D$ is related to the electric field $E$, polarization $P$, and permittivity $\varepsilon_0$ of free-space via $D = \varepsilon_0 E + P$, and magnetic induction $B$ is related to the magnetic field $H$, magnetization $M$, and permeability $\mu_0$ of free-space via $B = \mu_0 H + M$. These equations tie the electromagnetic (EM) fields $E(r,t)$ and $H(r,t)$ to their sources $\rho_{\text{free}}, J_{\text{free}}, P,$ and $M$, which generally have arbitrary distributions throughout space-time $(r,t)$.

Maxwell's equations are silent as to the energy content of the EM fields. However, once the Poynting vector $S(r,t) = E(r,t) \times H(r,t)$ is defined as the universal expression of the flow-rate of EM energy, Eqs.(1)-(4) lead straightforwardly to Poynting's theorem, that is,

$$\nabla \cdot S + \frac{\partial}{\partial t}(\tfrac{1}{2}\varepsilon_0 E \cdot E + \tfrac{1}{2}\mu_0 H \cdot H)$$
$$+ \left(E \cdot J_{\text{free}} + E \cdot \tfrac{\partial P}{\partial t} + H \cdot \tfrac{\partial M}{\partial t}\right) = 0. \quad (5)$$

Not only does the above equation yield expressions for the field energy densities ($\tfrac{1}{2}\varepsilon_0 E^2$ and $\tfrac{1}{2}\mu_0 H^2$) as well as the energy-exchange-rate between material media and the $E$ and $H$ fields [that is, $E \cdot (J_{\text{free}} + \partial P/\partial t)$ and $H \cdot \partial M/\partial t$], but also its form (as a continuity equation) *guarantees* the conservation of EM energy.

Similarly, EM force and momentum do not emerge naturally from Maxwell's equations. However, once a stress tensor is defined, Eqs.(1)-(4) can be invoked to arrive at expressions for the densities of EM force, torque, momentum, and angular momentum.[1,2] As an example, let us consider the Einstein-Laub stress tensor

$$\overleftrightarrow{\mathcal{T}}_{EL}(r,t) = \tfrac{1}{2}(\varepsilon_0 E \cdot E + \mu_0 H \cdot H)\overleftrightarrow{\mathbf{I}} - DE - BH, \quad (6)$$

where $\overleftrightarrow{\mathbf{I}}$ is the 2nd rank identity tensor—a $3 \times 3$ matrix.[3] This stress tensor can be readily shown to satisfy the continuity equation[4]

$$\nabla \cdot \overleftrightarrow{\mathcal{T}}_{EL} + \frac{\partial}{\partial t}(E \times H/c^2) + F_{EL}(r,t) = 0, \quad (7)$$

where the Einstein-Laub force-density $F_{EL}$ is given by

$$F_{EL}(r,t) = \rho_{\text{free}} E + J_{\text{free}} \times \mu_0 H + (P \cdot \nabla)E$$
$$+ \tfrac{\partial P}{\partial t} \times \mu_0 H + (M \cdot \nabla)H - \tfrac{\partial M}{\partial t} \times \varepsilon_0 E. \quad (8)$$

The formulation proposed by Einstein and Laub thus assigns to the EM field the momentum-density $\wp_{EL}(r,t) = S(r,t)/c^2$, which has been associated with the name of Max Abraham.[5] The Einstein-Laub force-density $F_{EL}$ is the force exerted by the $E$ and $H$ fields on material media, which are the seats of $\rho_{\text{free}}, J_{\text{free}}, P,$ and $M$. Considering that "force" is the rate of transfer of mechanical momentum to (or from) material media, the continuity equation (7) *guarantees* the conservation of linear momentum.

If the position-vector $r$ is cross-multiplied into Eq.(7), we arrive[4] at a similar equation for the conservation of angular momentum, where the EM angular momentum density will be given by $\mathcal{L}_{EL}(r,t) = r \times S(r,t)/c^2$ and the EM torque-density will be

$$T_{EL}(r,t) = r \times F_{EL} + P \times E + M \times H. \quad (9)$$

It must be emphasized that the above discussion is completely general, depending in no way on the nature of the sources. The material media hosting the sources could respond linearly or nonlinearly to the EM fields, they could be mobile or stationary, they could have permanent polarization and magnetization, or their $P$ and $M$ could be induced by local or non-local fields, etc. Maxwell's macroscopic equations and the stress tensor formulation of force, torque, and momentum are applicable under all circumstances.[6]

There exist several other stress tensors, each with its own expressions for the densities of EM force, torque, momentum, and angular momentum.[1,5] For instance, in the standard (Lorentz) formulation of classical electrodynamics, the stress tensor is defined as

$$\overleftrightarrow{\mathcal{T}}_L(r,t) = \tfrac{1}{2}(\varepsilon_0 E \cdot E + \mu_0^{-1} B \cdot B)\overleftrightarrow{\mathbf{I}} - \varepsilon_0 EE - \mu_0^{-1} BB, \quad (10)$$

whereas Minkowski's stress tensor is

$$\overleftrightarrow{\mathcal{T}}_M(r,t) = \tfrac{1}{2}(D \cdot E + B \cdot H)\overleftrightarrow{\mathbf{I}} - DE - BH. \quad (11)$$

Each formulation, of course, complies with the conservation laws, provided that the relevant entities are properly defined in accordance with continuity equations such as Eqs.(5) and (7). Occasional confusions in the literature can be traced to the fact that some authors use, for instance, the momentum-density from one formulation and the force-density from another. Needless to say, if the various entities are used properly (i.e., in the context of a single stress tensor), there should be no confusion and no inconsistency. That is *not* to say that all the existing formulations are equivalent. There are theoretical arguments (e.g., the Balazs thought experiment[7]), which reveal that certain formulations violate well-established physical principles.[8] Other formulations might require the introduction of so-called "hidden" entities such as hidden energy and hidden momentum.[4,9,10] Of course, the ultimate proof of validity of a physical theory is its compliance with experimental observations. The debate as to which stress tensor represents physical reality is very much alive today, and the interested reader should consult the vast literature of the subject.[1,2,5,10-15] The present author believes that several appealing features of the Einstein-Laub theory recommend it as a universal theory of EM force and momentum,[4,6,16] but there are others who prefer alternative formulations.[5,10,11]

The following example illustrates the type of results that one can obtain by a consistent application of a specific stress tensor to a given physical problem.

**Example**. Figure 1 shows a light pulse of frequency $\omega$ entering a transparent slab of material identified by its permittivity $\varepsilon(\omega)$ and permeability $\mu(\omega)$. The slab's refractive index and impedance are $n = \sqrt{\mu\varepsilon}$ and $\eta = \sqrt{\mu/\varepsilon}$; the Fresnel reflection coefficient at the entrance facet of the slab is $\rho = (\eta - 1)/(\eta + 1)$.

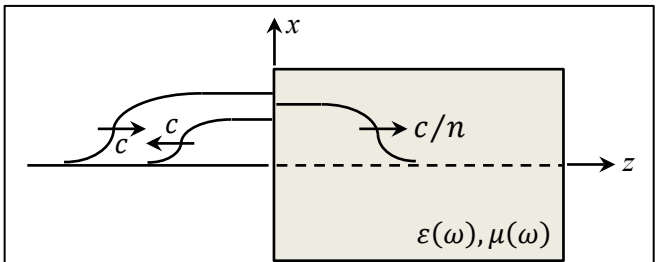

Figure 1. Linearly polarized light pulse of frequency $\omega$ entering a homogeneous, isotropic slab identified by its permittivity $\varepsilon(\omega)$ and permeability $\mu(\omega)$. The incidence medium is free space, in which the incident and reflected pulses propagate at the vacuum speed of light, $c$. Inside the slab, the pulse propagates at the reduced velocity $c/n$, where the refractive index is given by $n = \sqrt{\mu\varepsilon}$.

The incident beam, whose cross-sectional area is $A$, is linearly polarized along the $x$-axis, its $\boldsymbol{E}$ and $\boldsymbol{H}$ fields being $E_0\hat{\boldsymbol{x}}\cos(k_0 z - \omega t)$ and $H_0\hat{\boldsymbol{y}}\cos(k_0 z - \omega t)$, where $H_0 = E_0/Z_0$ and $k_0 = \omega/c$; here $Z_0 = \sqrt{\mu_0/\varepsilon_0}$ is the impedance of free space. Considering that the time-rate of arrival of EM momentum at the entrance facet is $\tfrac{1}{2}\varepsilon_0 E_0^2 A$, the rate at which EM momentum is delivered to the slab is $\tfrac{1}{2}(1 + \rho^2)\varepsilon_0 E_0^2 A$.

Immediately after the entrance facet, the field amplitudes entering the slab are $E_1 = (1 + \rho)E_0$ and $H_1 = (1 - \rho)H_0$, yielding the corresponding EM momentum-density $\wp_{EL} = \tfrac{1}{2}(1 - \rho^2)\varepsilon_0 E_0^2/c$. Given that the velocity of the leading edge of the pulse within the slab is $c/n$, the (Abraham) momentum-content of the slab increases at the rate of $\tfrac{1}{2}(1 - \rho^2)\varepsilon_0 E_0^2 A/\sqrt{\mu\varepsilon}$. The difference between the rate of delivery of EM momentum from the outside and the growth rate of EM momentum within the slab, namely, $(\varepsilon + \mu - 2)\varepsilon_0 E_0^2/(\sqrt{\varepsilon} + \sqrt{\mu})^2$, must be taken up by the EM force exerted on the slab at the leading edge of the light pulse. This is indeed the case, as can be readily verified by integrating the force-density $\boldsymbol{F}_{EL}(\boldsymbol{r}, t)$ of Eq.(8) over the volume of the slab. Calculation of this force requires defining the function $f(z)$ to represent the fields inside the slab as $\boldsymbol{E}(z,t) = E_1\hat{\boldsymbol{x}}f(z - ct/n)$ and $\boldsymbol{H}(z,t) = H_1\hat{\boldsymbol{y}}f(z - ct/n)$; material polarization and magnetization now become $\boldsymbol{P}(z,t) = \varepsilon_0(\varepsilon - 1)\boldsymbol{E}(z,t)$ and $\boldsymbol{M}(z,t) = \mu_0(\mu - 1)\boldsymbol{H}(z,t)$. (Note that material dispersion is ignored.) Only the 4th and 6th terms on the right-hand side of Eq.(8) contribute to the force-density in the present example. The requisite calculations are fairly elementary and need not be elaborated here.

Finally, if the pulse duration is denoted by $\tau$, the total energy and momentum delivered to the slab, namely,

$$\mathcal{E} = \tfrac{1}{2}(1 - \rho^2)Z_0^{-1}E_0^2 A\tau, \qquad (12)$$

$$\boldsymbol{\wp} = \tfrac{1}{2}(1 + \rho^2)\varepsilon_0 E_0^2 A\tau\hat{\boldsymbol{z}}, \qquad (13)$$

yield $\wp/\mathcal{E} = \tfrac{1}{2}\sqrt{\mu_0\varepsilon_0}(\sqrt{\mu/\varepsilon} + \sqrt{\varepsilon/\mu})$. Thus, for a single photon of energy $\hbar\omega$ traveling in the slab, the total (i.e., electromagnetic + mechanical) momentum is bound to be $(\sqrt{\mu/\varepsilon} + \sqrt{\varepsilon/\mu})\hbar\omega/2c$.[6]

**Current and Future Challenges**. On the fundamental side, there is a need to clarify the differences among the various stress tensors, namely, those of Lorentz, Minkowski, Abraham, Einstein-Laub, and Chu.[1,5,11] Contrary to a widely-held belief, these tensors *can* be distinguished from each other by the subtle differences in their predicted *distributions* of force- and/or torque-density in deformable media.[16] Experimental tests on transparent (i.e., non-absorbing) magnetic materials are particularly welcome in this regard, as some of the major differences among the proposed stress tensors emerge in magnetized or magnetisable media.[4,5,15]

**Advances in Science and Technology to Meet Challenges**. The observable effects of EM force and torque on material media are enhanced when the fields are made to interact with micro- and nano-structured objects, especially those possessing extremely large quality-factors — also known as high-Q resonators. Quantum opto-mechanics has blossomed in recent years as a result of advances in micro/nano-fabrication. These advances are likely to continue as the fabrication tools and measurement techniques improve, and as novel meta-materials are discovered.

**Concluding Remarks**. The classical theory of electrodynamics suffers from an embarrassment of riches in that several stress tensors have been proposed over the years that provide alternative expressions for EM force, torque, and momentum in material media.[4,5] Advances in meta-materials, micro-structured material fabrication, and measurement tools and techniques have reached a stage where it is now possible to expect that experiments in the near future will be able to distinguish among the extant EM stress tensors.

**References**

[1] P. Penfield and H. A. Haus, *Electrodynamics of Moving Media*, MIT Press, Cambridge, MA (1967).
[2] F. N. H. Robinson, "Electromagnetic stress and momentum in matter," *Phys. Rep.* **16**, 313-354 (1975).

[3] A. Einstein and J. Laub, "Über die im elektromagnetischen Felde auf ruhende Körper ausgeübten ponderomotorischen Kräfte," *Ann. Phys.* **26**, 541-550 (1908); reprinted in translation: "On the ponderomotive forces exerted on bodies at rest in the electromagnetic field," *The Collected Papers of Albert Einstein*, vol.2, Princeton University Press, Princeton, NJ (1989).

[4] M. Mansuripur, "The Force Law of Classical Electrodynamics: Lorentz versus Einstein and Laub," Optical Trapping and Optical Micro-manipulation X, edited by K. Dholakia and G. C. Spalding, *Proc. of SPIE* **8810**, 88100K~1:18 (2013).

[5] B. A. Kemp, "Resolution of the Abraham-Minkowski debate: Implications for the electromagnetic wave theory of light in matter," *J. Appl. Phys.* **109**, 111101 (2011).

[6] M. Mansuripur, "On the Foundational Equations of the Classical Theory of Electrodynamics," *Resonance* **18**(2), 130-155 (2013).

[7] N. L. Balazs, "The energy-momentum tensor of the electromagnetic field inside matter," *Phys. Rev.* **91**, 408-411 (1953).

[8] R. Loudon, "Radiation pressure and momentum in dielectrics," *Fortschr. Phys.* **52**, 1134-1140 (2004).

[9] W. Shockley and R. P. James, "'Try simplest cases' discovery of 'hidden momentum' forces on 'magnetic currents,'" *Phys. Rev. Lett.* **18**, 876-879 (1967).

[10] D. J. Griffiths and V. Hnizdo, "Mansuripur's Paradox," *Am. J. Phys.* **81**, 570-573 (2013).

[11] I. Brevik, "Experiments in phenomenological electrodynamics and the electromagnetic energy-momentum tensor," *Phys. Rep.* **52**, 133-201 (1979).

[12] R. Loudon, "Theory of the radiation pressure on dielectric surfaces," *J. Mod. Opt.* **49**, 821-38 (2002).

[13] C. Baxter and R. Loudon, "Radiation pressure and the photon momentum in dielectrics," *J. Mod. Opt.* **57**, 830-842 (2010).

[14] S. M. Barnett and R. Loudon, "The enigma of optical momentum in a medium," *Phil. Trans. R. Soc. A* **368**, 927-939 (2010).

[15] S. M. Barnett and R. Loudon, "Theory of radiation pressure on magneto-dielectric materials," *New J. Phys.* **17**, 063027, pp 1-16 (2015).

[16] M. Mansuripur, A.R. Zakharian, and E.M. Wright, "Electromagnetic-force distribution inside matter," *Phys. Rev. A* **88**, 023826, pp1-13 (2013).